\documentclass[journal]{IEEEtran}
\usepackage{graphicx}

\usepackage{bm}
\usepackage{amsfonts}

\usepackage{amssymb}

\usepackage{color}

\usepackage{cite}

\ifCLASSINFOpdf
\else
\fi
\usepackage{amsmath}
\usepackage{algorithm}
\usepackage{algorithmic}
\ifCLASSOPTIONcompsoc
 \usepackage[caption=false,font=normalsize,labelfont=sf,textfont=sf]{subfig}
\else
 \usepackage[caption=false,font=footnotesize]{subfig}
\fi
\begin{document}
%
\title{An Efficient Manifold Algorithm for Constructive Interference based Constant Envelope Precoding}
%
%
%

\author{Fan Liu,~\IEEEmembership{Student~Member,~IEEE,}
        Christos Masouros,~\IEEEmembership{Senior~Member,~IEEE,}
        Pierluigi Vito Amadori,~\IEEEmembership{Student~Member,~IEEE,}
        and~Huafei Sun
\thanks{Manuscript received ***. This work was supported by the Engineering and Physical
Sciences Research Council (EPSRC) project EP/M014150/1 and the China
Scholarship Council (CSC).}
\thanks{F. Liu is with the School of Information and Electronics, Beijing Institute of Technology, Beijing, 100081, China, and is also with the Department of Electronic and Electrical Engineering, University College London, London, WC1E 7JE, UK (e-mail: liufan92@bit.edu.cn).} 
\thanks{C. Masouros and P. V. Amadori are with the Department of Electronic and Electrical Engineering, University College London, London, WC1E 7JE, UK (e-mail: chris.masouros@ieee.org, uceepva@ucl.ac.uk).}
\thanks{H. Sun is with the School of Mathematics and Statistics, Beijing Institute of Technology, Beijing, 100081, China (email: huafeisun@bit.edu.cn).}
}

\maketitle

\begin{abstract}
In this letter, we propose a novel manifold-based algorithm to solve the constant envelope (CE) precoding problem with interference exploitation. For a given power budget, we design the precoded symbols subject to the CE constraints, such that the constructive effect of the multi-user interference (MUI) is maximized. While the objective for the original problem is non-differentiable on the complex plane, we consider the smooth approximation of its real representation, and map it onto a Riemannian manifold. By using the Riemmanian conjugate gradient (RCG) algorithm, a local minimizer can be efficiently found for the problem. The complexity of the algorithm is analytically derived in terms of floating-points operations (flops) per iteration. Numerical results show that the proposed algorithm outperforms the conventional methods on both symbol error rate and computational complexity. 
\end{abstract}

\begin{IEEEkeywords}
Constant envelope, MU-MISO downlink, massive MIMO, manifold optimization.
\end{IEEEkeywords}

%
\IEEEpeerreviewmaketitle

\section{Introduction}
%
%
%
%
\IEEEPARstart{A}{S} one of the most promising approaches in 5G technology, massive multi-input-multi-output (mMIMO) communication systems are expected to provide significant benefits over conventional MIMO systems by employing much larger antenna arrays\cite{6736761,5595728}. Nevertheless, such systems face numerous challenges brought by the increasing number of antennas, e.g., higher hardware costs and power consumption, which may delay its deployment in future 5G systems. Hence, cheap and efficient RF power amplifiers (PA) are required for making the technology realizable in practical scenarios. 
\\\indent It is important to note that most of power-efficient PAs are made by non-linear components, therefore waveforms with low peak-to-average-power-ratio (PAPR) are needed to avoid signal distortions when the PA is operated at the saturation region\cite{5978417}. Pioneered by \cite{6297982,6451071}, the constant envelope precoding (CEP) has been proposed as an enabling solution, where the MUI is minimized subject to the CE constraints. The optimization in \cite{6451071} is a non-convex non-linear least square (NLS) problem, and is solved by sequential gradient descent (GD) method, which converges to a local minimum. To further improve the performance, a cross-entropy optimization (CEO) solver is introduced in \cite{6853373}. More recently, by using the fact that the feasible region of the CE problem can be geometrically viewed as a complex circle manifold, a RCG algorithm is proposed by \cite{7811286}, where the NLS problem is solved with much lower complexity than both GD and CEO. While the interference reduction (IR) methods in above works are relatively straightforward, their performance is strongly dependent with the constellation energy \cite{6451071}, which is difficult to optimally set in advance. In addition, IR approaches ignore that MUI is known to the base station (BS) in general, and thus can be utilized as a source of useful power. Realizing these facts, the previous work \cite{7738555} considers a novel CEP approach with the concept of constructive interference (CI) \cite{7103338}, which can overcome the above drawbacks. Due to the CE constraints, the CI-CEP problem is non-convex, but can be solved using CEO solver as well. Moreover, by relaxing the constraints, the CI-CEP problem becomes convex, thus can be solved by standard numerical tools. However, both of the above methods demand large amount of computations inevitably. 
\\\indent Based on the previous works on manifold optimizations \cite{duan2013natural,li2016optimal}, we consider a manifold-based algorithm to solve the CI-CEP problem in this letter. Since the objective is not complex differentiable, we first equivalently transform the problem into its real representation, and use a smooth upper-bound to obtain a differentiable approximation. By viewing the feasible region as an oblique manifold, a RCG algorithm is employed to find a local minimizer of the problem. Unlike the relaxed convex problem in \cite{7738555}, the proposed algorithm is guaranteed to yield precoded symbols with exactly constant envelopes, and has better performance than the methods of \cite{7738555} in terms of both symbol error rate (SER) and complexity. 

\section{System Model}
We consider a multi-user multi-input-single-output (MU-MISO) downlink scenario where a \emph{N}-antenna BS transmits signals to \emph{M} single-antenna users. The received signal vector is given as
\begin{equation}
{\mathbf{y}} = {{\mathbf{H}}^T\mathbf{x}} + {\mathbf{w}},
\end{equation}
where ${\mathbf{y}} = {\left[ {{y_1},{y_2},...,{y_M}} \right]^T} \in {\mathbb{C}^{M \times 1}}$ with $y_m$ being the received symbol for the \emph{m}-th user, ${\mathbf{x}} = {\left[ {{x_1},{x_2},...,{x_N}} \right]^T} \in {\mathbb{C}^{N \times 1}}$ represents the transmitted symbols, ${\mathbf{w}} = {\left[ {{w_1},{w_2},...,{w_M}} \right]^T} \in {\mathbb{C}^{M \times 1}} \sim \mathcal{C}\mathcal{N}\left( {0,{N_0}\mathbf{I}} \right)$ is the Gaussian noise, and ${\mathbf{H}} = \left[ {{{\mathbf{h}}_1},{{\mathbf{h}}_2},...,{{\mathbf{h}}_M}} \right] \in {\mathbb{C}^{N \times M}}$ is the channel matrix, with $\mathbf{h}_m$ being the channel vector for the \emph{m}-th user. Without loss of generality, the channel is assumed to be Rayleigh fading, i.e., each entry of $\mathbf{H}$ subjects to i.i.d Complex Gaussian distribution with zero-mean, and is perfectly known to the BS. The transmitted signal is expected to have constant envelope, which is
\begin{equation}
x_n = \sqrt{P_T/N}e^{j\theta_n}, \forall n,
\end{equation}
where $P_T$ is the total transmit power, $\theta_n$ is the phase of the \emph{n}-th transmitted symbol.
\\\indent Assume that the desired symbol for the \emph{m}-th user is $s_m = \sqrt{E_m}e^{j\phi_m}$, where $E_m$ and $\phi_m$ denote the power and the phase of the symbol respectively. The received symbol for the \emph{m}-th user can be written as
\begin{equation}
{y_m} = {s_m} + \underbrace {\left( {{\mathbf{h}}_m^T{\mathbf{x}} - {s_m}} \right)}_{\text{MUI}} + {w_m},
\end{equation}
where the second term represents the interfering signal for the user. The total MUI power is then given by
\begin{equation}
{P_{MUI}} = \sum\limits_{m = 1}^M {{{\left( {{\mathbf{h}}_m^T{\mathbf{x}} - {s_m}} \right)}^2} = } {\left\| {{\mathbf{H}^T\mathbf{x}} - {\mathbf{s}}} \right\|^2},
\end{equation}
where $\mathbf{s}=\left[s_1,s_2,...,s_M\right]^T$ is the desired symbol vector.
\section{Problems Formulation}
Aiming at minimizing the MUI power, the conventional CEP approaches are designed to solve the following optimization problem \cite{6451071}
\begin{equation}
\begin{gathered}
  \mathop {\min }\limits_{\mathbf{x}} \;\;{\left\| {{{\mathbf{H}}^T}{\mathbf{x}} - {\mathbf{s}}} \right\|^2} \hfill \\
  s.t.\;\;\;\left| {{x_n}} \right| = \sqrt {P_T/N} ,\forall n. \hfill \\ 
\end{gathered}
\end{equation}
Problem (5) is a NLS problem, which is obviously non-convex, and has multiple local minima. Fortunately, it has been proven that most of the local minima yield small values\cite{6451071}, and can be obtained by a variety of approaches \cite{6451071,6853373,7811286}. However, it should be highlighted that by treating all the interference as harmful, these techniques ignore the fact that MUI can be employed as a green signal power source to benefit the symbol demodulation. This has been first proposed by \cite{4801492}, where the MUI is classified as constructive and destructive parts. CI based beamformers aim at minimizing destructive and exploiting constructive interference, which enable a relaxed feasible region for the optimization \cite{7103338}. Based on this, previous work \cite{7738555} focuses on maximizing the \emph{constructive effect} of the MUI to achieve CE precoding, where the PSK modulations are employed. We refer the reader to the above literature for detailed discussions. Here we recapture the CI-CEP problem in \cite{7738555} as follows
\begin{equation}
\begin{gathered}
  \mathop {\min }\limits_{\mathbf{x}} \mathop {\max }\limits_m \;\left| {\operatorname{Im} \left( {{t_m}} \right)} \right|\; - \operatorname{Re} \left( {{t_m}} \right)\tan \psi  \hfill \\
  s.t.\;\;\;\left| {{x_n}} \right| = \sqrt {{P_T}/N} ,\forall n,\;\;\;\;\;\;\;\;\;\;\;\;\;\;\; \hfill \\
  \;\;\;\;\;\;\;\;\;{t_m} = \left( {{\mathbf{h}}_m^T{\mathbf{x}} - {s_m}} \right){e^{ - j{\phi _m}}},\forall m, \hfill \\ 
\end{gathered} 
\end{equation}
where $s_m = ue^{j\phi_m}$, $\psi = \pi/L$, $u$ is the amplitude for the PSK symbols, $L$ is the PSK modulation order. The above problem can be solved by CEO suboptimally, and has been further relaxed as a convex problem by replacing the equality constraints on $x_n$ as inequalities, i.e., $\left| {{x_n}} \right| \le \sqrt {{P_T}/N} ,\forall n$. Such a convex approximation problem can be efficiently solved by numerical solvers, e.g., CVX toolbox. The results are then normalized to obtain transmitted symbols with constant envelopes \cite{7738555}. Nevertheless, using CEO or CVX to solve (6) requires significant computation resources. In the next section, we propose a manifold based optimization technique to solve (6), which has much lower complexity.  
\section{Proposed Algorithm based on Oblique Manifold}
Since $\operatorname{Re}\left(\cdot\right)$ and $\operatorname{Im}\left(\cdot\right)$ are not complex differentiable, we formulate the real representation of (6). First we rewrite $t_m$ as
\begin{equation}
{t_m} = \left( {{\mathbf{h}}_m^T{\mathbf{x}} - {s_m}} \right){e^{ - j{\phi _m}}} = {\mathbf{\tilde h}}_m^T{\mathbf{x}} - u,
\end{equation}
where ${\mathbf{\tilde h}}_m={\mathbf{h}}_m{e^{ - j{\phi _m}}}$. We then separate the real and imaginary parts of complex notations as follows
\begin{equation}
    {\mathbf{\tilde H}} = {{{\mathbf{\tilde H}}}_R} + j{{{\mathbf{\tilde H}}}_I},{{{\mathbf{\tilde h}}}_m} = {{{\mathbf{\tilde h}}}_{Rm}} + j{{{\mathbf{\tilde h}}}_{Im}},{\mathbf{x}} = {{\mathbf{x}}_R} + j{{\mathbf{x}}_I}.
\end{equation}
where ${\mathbf{\tilde H}} = \left[ {{{{\mathbf{\tilde h}}}_1},{{{\mathbf{\tilde h}}}_2},...,{{{\mathbf{\tilde h}}}_M}} \right]$. It follows that
\begin{equation}
\begin{gathered}
  \operatorname{Re} \left( {{t_m}} \right) = {\mathbf{\tilde h}}_{Rm}^T{{\mathbf{x}}_R} - {\mathbf{\tilde h}}_{Im}^T{{\mathbf{x}}_I} - u, \hfill \\
  \operatorname{Im} \left( {{t_m}} \right) = {\mathbf{\tilde h}}_{Im}^T{{\mathbf{x}}_R} + {\mathbf{\tilde h}}_{Rm}^T{{\mathbf{x}}_I}. \hfill \\ 
\end{gathered} 
\end{equation}
By using the fact that $\left| a \right| = \max \left( {a, - a} \right)$, and denoting $\beta=\tan \psi$ we have
\begin{equation}
    \left| {\operatorname{Im} \left( {{t_m}} \right)} \right| - \operatorname{Re} \left( {{t_m}} \right)\tan \psi  = \max \left( {{g_{2m-1}},{g_{2m}}} \right) - u\beta,
\end{equation}
where
\begin{equation}
\begin{gathered}
  {g_{2m-1}} = {\left( {{{{\mathbf{\tilde h}}}_{Im}} - \beta{{{\mathbf{\tilde h}}}_{Rm}} } \right)^T}{{\mathbf{x}}_R} + {\left( {{{{\mathbf{\tilde h}}}_{Rm}} + \beta{{{\mathbf{\tilde h}}}_{Im}} } \right)^T}{\mathbf{x}}_I, \hfill \\
  {g_{2m}} = {\left( \beta{{{{\mathbf{\tilde h}}}_{Im}}  - {{{\mathbf{\tilde h}}}_{Rm}}} \right)^T}{{\mathbf{x}}_I} - {\left( {{{{\mathbf{\tilde h}}}_{Im}} + \beta{{{\mathbf{\tilde h}}}_{Rm}} } \right)^T}{{\mathbf{x}}_R}. \hfill \\ 
\end{gathered} 
\end{equation}
Denoting ${\mathbf{\tilde X}} = \sqrt{\frac{N}{P_T}}{\left[ {{{\mathbf{x}}_R},{{\mathbf{x}}_I}} \right]^T}$, the real representaion of the problem can be written compactly as follows 
\begin{equation}
\begin{gathered}
  \mathop {\min }\limits_{{\mathbf{\tilde X}}} \mathop {\max }\limits_i {g_i} \hfill \\
  s.t.\;\;{\left( {{{{\mathbf{\tilde X}}}^T}{\mathbf{\tilde X}}} \right)_{nn}} = 1,n = 1,2,...,N, \hfill \\ 
\end{gathered} 
\end{equation}
where $i=1,2,...,2M$. It is clear that the feasible region of (12) can be given as
\begin{equation}
\mathcal{M} = \left\{ {{\mathbf{\tilde X}} \in {\mathbb{R}^{2 \times N}}:{{({{{\mathbf{\tilde X}}}^T}{\mathbf{\tilde X}})}_{nn}} = 1,\forall n} \right\}.
\end{equation}
We say that $\mathcal{M}$ forms a \emph{manifold}, and ${\mathbf{\tilde X}}$ is a point on $\mathcal{M}$. To be more specific, $\mathcal{M}$ is a $2N$-dimensional \emph{oblique manifold}\cite{boumal2014manopt}. In Riemannain geometry, a manifold is defined as a set of points that endowed with a locally Euclidean structure near each point. Given a point $p$ on $\mathcal{M}$, a \emph{tangent vector} at $p$ is defined as the vector that is tangent to any smooth curves on $\mathcal{M}$ through $p$. The set of all such vectors at $p$ forms the \emph{tangent space}, denoted by $T_p\mathcal{M}$, which is an Euclidean space. Specially, the tangent space at ${\mathbf{\tilde X}}$ is given as
\begin{equation}
{T_{{\mathbf{\tilde X}}}}\mathcal{M} = \left\{ {{\mathbf{U}} \in {\mathbb{R}^{2 \times N}}:{{({{{\mathbf{\tilde X}}}^T}{\mathbf{U}})}_{nn}} = 0,\forall n} \right\}.
\end{equation}
If the tangent spaces of a manifold are equipped with a smoothly varying inner product, the manifold is called \emph{Riemannian manifold}\cite{petersen2006riemannian}. Accordingly, the family of inner products is called \emph{Riemannian metric}, which allows the existence of rich geometric structure on the manifold. Here we use the usual Euclidean inner product as the metric, which is ${\left\langle {{\mathbf{U}},{\mathbf{V}}} \right\rangle _{{\mathbf{\tilde X}}}} = \operatorname{tr} \left( {{{\mathbf{U}}^T}{\mathbf{V}}} \right)$, where ${{\mathbf{U}},{\mathbf{V}}} \in {T_{{\mathbf{\tilde X}}}}\mathcal{M}$. 
\begin{figure}
    \centering
    \includegraphics[width=2.8in]{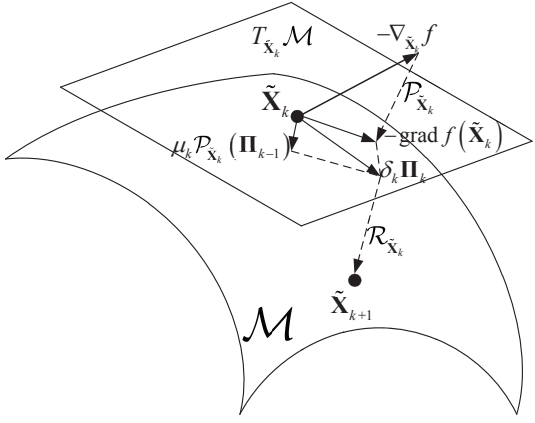}
    \caption{Riemannian conjugate gradient algorithm.}
    \label{fig:1}
\end{figure}
The algorithm that we employ is the so-called Riemannian conjugate gradient (RCG) algorithm \cite{boumal2014thesis}, which needs to first compute the gradient of the objective. Since the objective in (12) is still not differentiable, we consider the well-known smooth log-sum-exp upper-bound $f\left(\mathbf{\tilde X}\right)$ for the max function, which is
\begin{equation}
\begin{gathered}
  {g_{\max }} \le f\left( {\mathbf{\tilde X}} \right) = \varepsilon \log \left( {\sum\limits_i {\exp \left( {{g_i}/\varepsilon } \right)} } \right) \hfill \\
  \;\;\;\;\;\;\;\;\;\le {g_{\max }} + \varepsilon \log \left( {2M} \right), \hfill \\ 
\end{gathered}
\end{equation}
where $\varepsilon > 0$ is some small positive number. The gradient of $f\left( {\mathbf{\tilde X}} \right)$ is thus given as
\begin{equation}
{\nabla _{{\mathbf{\tilde X}}}}f = \left[ {\frac{{\partial f}}{{\partial {{{\mathbf{\tilde x}}}_1}}},\frac{{\partial f}}{{\partial {{{\mathbf{\tilde x}}}_2}}},...,\frac{{\partial f}}{{\partial {{{\mathbf{\tilde x}}}_N}}}} \right],
\end{equation}
where ${{{\mathbf{\tilde x}}}_n}$ is the \emph{n}-th column of ${{{\mathbf{\tilde X}}}}$. Noting that ${{\mathbf{x}}_R} = \sqrt {\frac{N}{{{P_T}}}} {{{\mathbf{\tilde X}}}^T}\left( {:,1} \right),{{\mathbf{x}}_I} = \sqrt {\frac{N}{{{P_T}}}} {{{\mathbf{\tilde X}}}^T}\left( {:,2} \right)$, which are the first and second column of ${{{\mathbf{\tilde X}}}^T}$ respectively, we have
\begin{equation}
\frac{{\partial {{\mathbf{x}}_R}}}{{\partial {{{\mathbf{\tilde x}}}_n}}} = \sqrt {\frac{N}{{{P_T}}}} {\left[ {{{\mathbf{e}}_n},{\mathbf{0}}} \right]},\frac{{\partial {{\mathbf{x}}_I}}}{{\partial {{{\mathbf{\tilde x}}}_n}}} = \sqrt {\frac{N}{{{P_T}}}} {\left[ {{\mathbf{0}},{{\mathbf{e}}_n}} \right]},
\end{equation}
where ${{\mathbf{e}}_n}\in\mathbb{R}^{N\times 1}$ have all-zero entries except that its \emph{n}-th entry equals to 1. Based on (17), the \emph{n}-th column of the gradient is given by
\begin{equation}
\frac{{\partial f}}{{\partial {{{\mathbf{\tilde x}}}_n}}} = \frac{{\displaystyle\sqrt {\frac{N}{{{P_T}}}} \sum\limits_{m = 1}^M {\left( {\left[ \begin{gathered}
  {a_{n,m}}, {b_{n,m}} \hfill \\
  {c_{n,m}},{d_{n,m}} \hfill \\ 
\end{gathered}  \right]\left[ \begin{gathered}
  \exp \left( {\frac{{{g_{2m - 1}}}}{\varepsilon }} \right) \hfill \\
  \exp \left( {\frac{{{g_{2m}}}}{\varepsilon }} \right) \hfill \\ 
\end{gathered}  \right]} \right)} }}{{\sum\limits_{i = 1}^{2M} {\exp \left( {\displaystyle\frac{{{g_i}}}{\varepsilon }} \right)} }},
\end{equation}
where ${a _{n,m}}, {b _{n,m}}, {c _{n,m}}$ and ${d_{n,m}}$ denote the $\left(n,m\right)$-th enrty of the following matrices
\begin{equation}
\begin{gathered}
  {\mathbf{A}} = {{{\mathbf{\tilde H}}}_I} - \beta {{{\mathbf{\tilde H}}}_R},{\mathbf{B}} =  - {{{\mathbf{\tilde H}}}_I} - \beta {{{\mathbf{\tilde H}}}_R}, \hfill \\
  {\mathbf{C}} = {{{\mathbf{\tilde H}}}_R}{\text{ + }}\beta {{{\mathbf{\tilde H}}}_I},{\mathbf{D}} = {{{\mathbf{\tilde H}}}_R} - \beta {{{\mathbf{\tilde H}}}_I}, \hfill \\ 
\end{gathered} 
\end{equation}
\\\indent In the RCG algorithm, (16) is called Euclidean gradient, and can be used to compute the Riemannian gradient, which is defined as the tangent vector belongs to $T_{{{\mathbf{\tilde X}}}}\mathcal{M}$ that indicates the steepest ascent direction of $f\left({{{\mathbf{\tilde X}}}}\right)$. It can be viewed as the orthogonal projection of the Euclidean gradient onto the tangent space\cite{absil2009optimization}, which is given as 
\begin{equation}
\operatorname{grad}f\left( {{\mathbf{\tilde X}}} \right)={\mathcal{P}_{{\mathbf{\tilde X}}}}\left( {{\nabla _{{\mathbf{\tilde X}}}}f} \right) = {\nabla _{{\mathbf{\tilde X}}}}f - {\mathbf{\tilde X}}\operatorname{diag} ({{{\mathbf{\tilde X}}}^T}{\nabla _{{\mathbf{\tilde X}}}}f),
\end{equation}
where ${\mathcal{P}_{{\mathbf{\tilde X}}}}\left(\cdot\right)$ denotes the projector, $\operatorname{diag}\left(\cdot\right)$ sets all off-diagonal entries of a matrix to zero. At the \emph{k}-th iteration, the descent direction $\mathbf {\Pi} _k$ is obtained as
\begin{equation}
\mathbf {\Pi}_k =  - \operatorname{grad} f\left( {{{{\mathbf{\tilde X}}}_k}} \right) + {\mu _k}{\mathcal{P}_{{{{\mathbf{\tilde X}}}_k}}}\left( {{\mathbf \Pi _{k - 1}}} \right).
\end{equation}
Here the projector is used as \emph{vector transport}, which maps the vector from one tangent space to another. $\mu_k$ is given by the Riemannian version of the Polak-Ribi\`ere formula, which is 
\begin{equation}
\begin{small}
\begin{gathered}
  {\mu _k} \hfill \\
   = \frac{{\left\langle {\operatorname{grad} f\left( {{{{\mathbf{\tilde X}}}_k}} \right),\operatorname{grad} f\left( {{{{\mathbf{\tilde X}}}_k}} \right) - {\mathcal{P}_{{{{\mathbf{\tilde X}}}_{k}}}}\left( {\operatorname{grad} f\left( {{{{\mathbf{\tilde X}}}_{k - 1}}} \right)} \right)} \right\rangle_{{{{\mathbf{\tilde X}}}_k}} }}{{\left\langle {\operatorname{grad} f\left( {{{{\mathbf{\tilde X}}}_{k - 1}}} \right),\operatorname{grad} f\left( {{{{\mathbf{\tilde X}}}_{k - 1}}} \right)} \right\rangle_{{{{\mathbf{\tilde X}}}_{k-1}}} }}. \hfill \\ 
\end{gathered}
\end{small}
\end{equation}
The \emph{k}$+1$-th update is thus given by
\begin{equation}
{{{\mathbf{\tilde X}}}_{k + 1}} = {\mathcal{R}_{{{{\mathbf{\tilde X}}}_k}}}\left( {{\delta _k}{{\mathbf{\Pi }}_k}} \right),
\end{equation}
where $ {\mathcal{R}_{{{{\mathbf{\tilde X}}}_k}}}\left(\cdot\right)$ is called \emph{retraction}, which maps a point on $T_{{{\mathbf{\tilde X}}}_k}\mathcal{M}$ to $\mathcal{M}$ with a local rigidity condition that preserves gradients at ${{{\mathbf{\tilde X}}}_k}$ \cite{absil2009optimization}, and is given as
\begin{equation}
\begin{gathered}
  {\mathcal{R}_{{{{\mathbf{\tilde X}}}_k}}}\left( {{\delta _k}{{\mathbf{\Pi }}_k}} \right) \hfill \\
   = \left[ {\frac{{{{\left( {{{{\mathbf{\tilde X}}}_k} + {\delta _k}{{\mathbf{\Pi }}_k}} \right)}_1}}}{{\left\| {{{\left( {{{{\mathbf{\tilde X}}}_k} + {\delta _k}{{\mathbf{\Pi }}_k}} \right)}_1}} \right\|}},...,\frac{{{{\left( {{{{\mathbf{\tilde X}}}_k} + {\delta _k}{{\mathbf{\Pi }}_k}} \right)}_N}}}{{\left\| {{{\left( {{{{\mathbf{\tilde X}}}_k} + {\delta _k}{{\mathbf{\Pi }}_k}} \right)}_N}} \right\|}}} \right], \hfill \\ 
\end{gathered} 
\end{equation}
where ${{{\left( {{{{\mathbf{\tilde X}}}_k} + {\delta _k}{\mathbf{\Pi } _k}} \right)}_n}}$ is the \emph{n}-th column of the matrix ${{\left( {{{{\mathbf{\tilde X}}}_k} + {\delta _k}{\mathbf{\Pi } _k}} \right)}}$, and the stepsize $\delta_k$ is obtained by backtracking line search algorithms, e.g., Armijo rule. Fig. 1 shows a single iteration of the RCG algorithm on $\mathcal{M}$, which has also been summarized in Algorithm 1.
\renewcommand{\algorithmicrequire}{\textbf{Input:}}
\renewcommand{\algorithmicensure}{\textbf{Output:}}
\begin{algorithm}
\caption{RCG for CI-based CEP}
\label{alg:A}
\begin{algorithmic}
    \REQUIRE $\mathbf{s}, \mathbf{H}, \Delta >0, k_{max}>0$.
    \ENSURE Local minimizer $\tilde{\mathbf{X}}^*$ for (12).
    \STATE 1. Initialize randomly $\tilde{\mathbf{X}}_0 \in \mathcal{M}$, \\ set $\mathbf{\Pi }_0=-\operatorname{grad}f\left(\tilde{\mathbf{X}}_0\right), k=0$,
    \WHILE{$k\le k_{max}$ \& ${\left\| {\operatorname{grad} f\left( {{{{\mathbf{\tilde X}}}_{k}}} \right)} \right\|_F} \ge {\Delta}$}
    \STATE 2. $k=k+1$,
    \STATE 3. Compute stepsize $\delta_{k-1}$ by Armijo rule, and set ${{{\mathbf{\tilde X}}}_{k}}$ using the retraction defined in (23),
    \STATE 4. Compute $\mu_k$ by (22),
    \STATE 5. Compute ${\mathbf{\Pi } _k}$ by (21).
    \ENDWHILE
\end{algorithmic}
\end{algorithm}
\begin{figure*}[!t]
\centering
\subfloat[]{\includegraphics[width=2.0in]{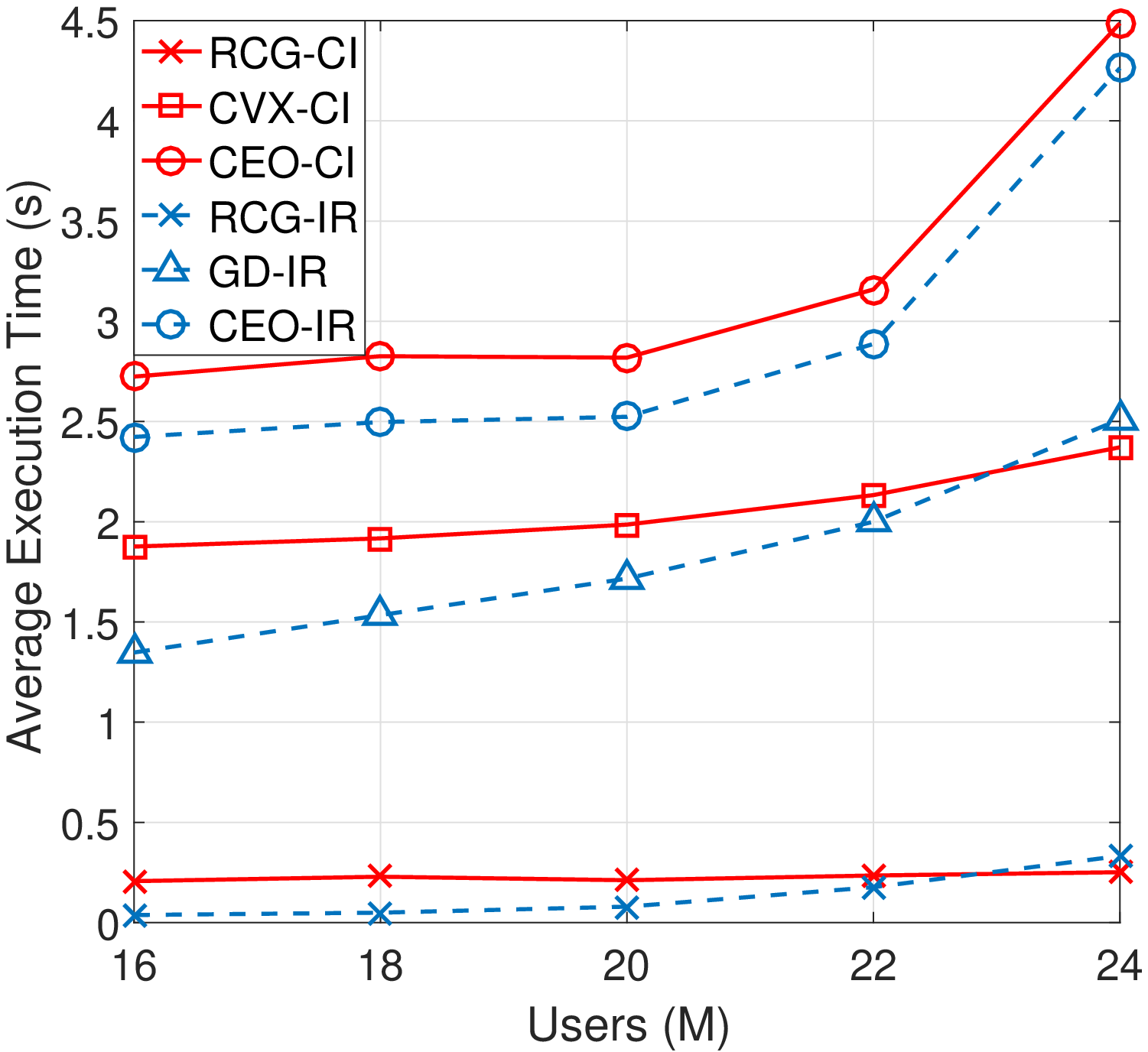}
\label{fig.1}}
\hspace{.01in}
\subfloat[]{\includegraphics[width=2.0in]{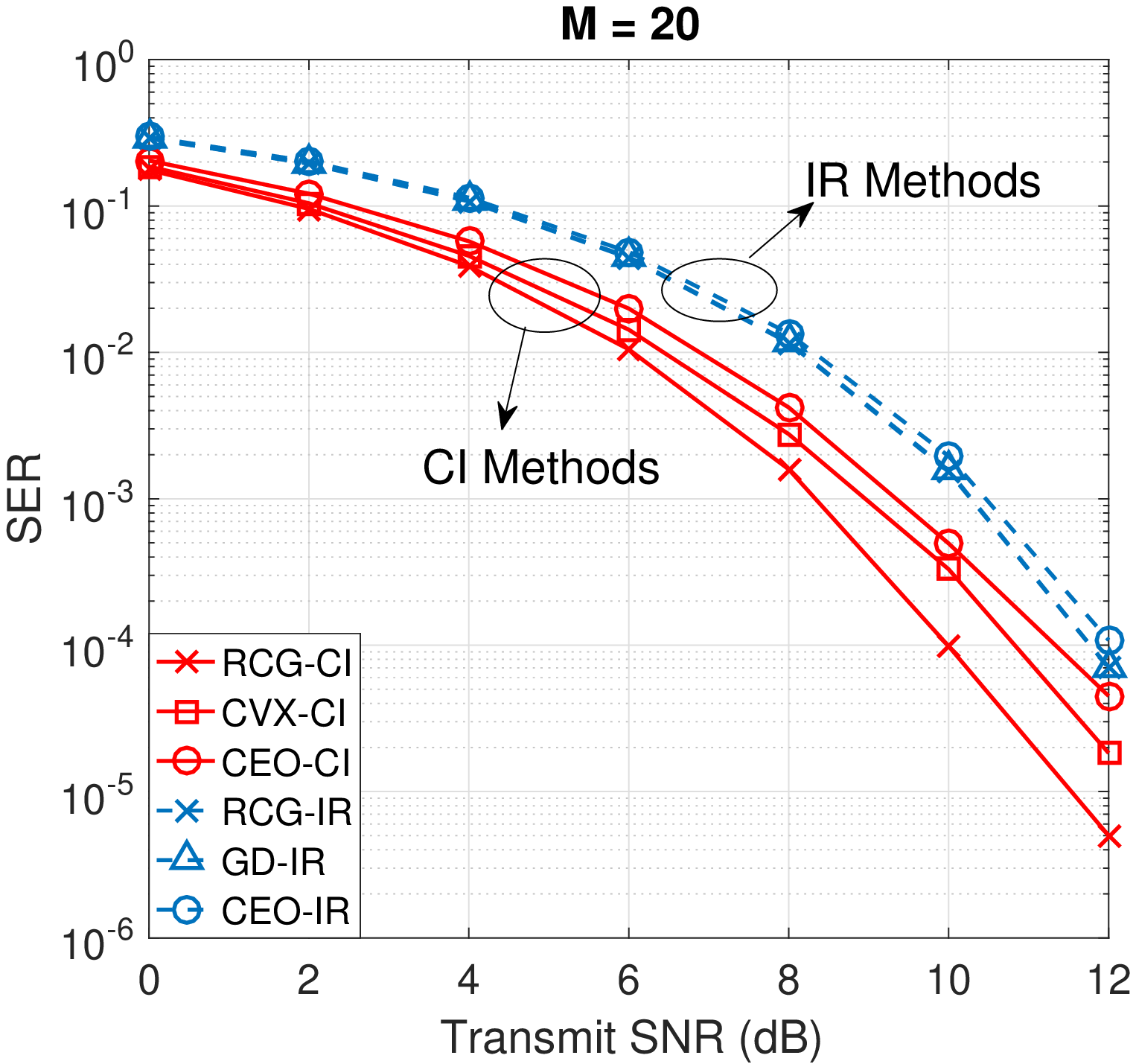}
\label{fig.2}}
\hspace{.01in}
\subfloat[]{\includegraphics[width=2.0in]{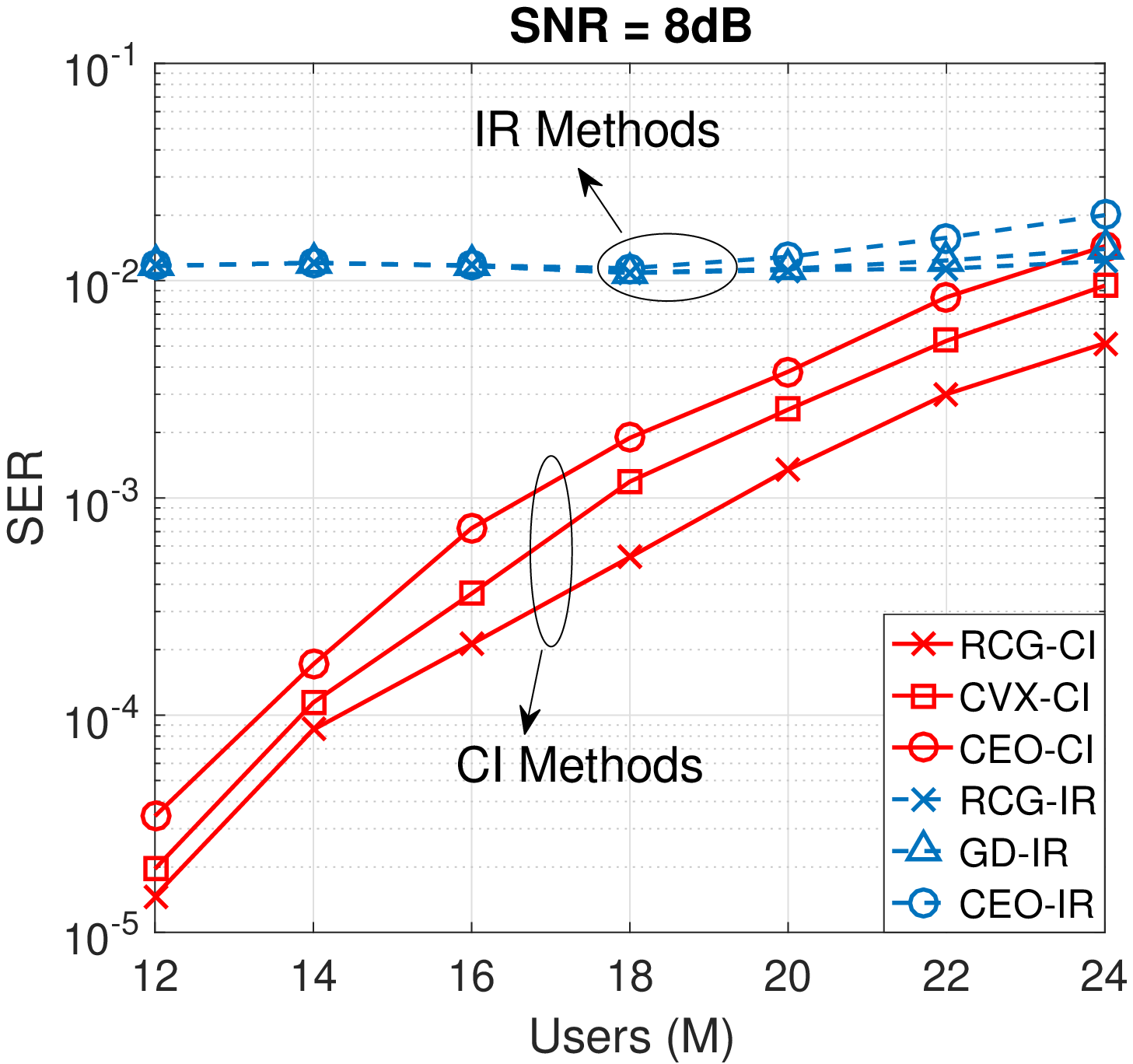}
\label{fig.3}}
\caption{Numerical results. (a) Average execution time vs. number of users for different algorithms; (b) SER vs. SNR for different algorithms; (c) SER vs. User for different algorithms.}
\label{fig_sim}
\end{figure*}

Note that the complexity of Algorithm 1 mainly comes from line 4 and line 5, where $16N^2+14MN+18M+16N$ and $4N^2+6N$ flops are required respectively, leading to a total complexity of $\mathcal{O}\left(N^2\right)$ for each iteration. By contrast, the complexity of GD and RCG-IR are $\mathcal{O}\left(MN^2\right)$ and $\mathcal{O}\left(MN\right)$ per iteration \cite{6451071,7811286}, respectively. For CEO, the complexity is $\mathcal{O}\left(KMN\right)$ in each iteration \cite{7738555}, where $K$ stands for the number of random samples, which may be quite larger than $M$ and $N$. While the RCG-IR requires less computations compared to the proposed algorithm, the latter brings significant performance gain as we will show in the next section, and therefore offers a favourable performance-complexity tradeoff.

\section{Numerical Results}
In this section, numerical results based on Monte Carlo simulations have been provided to compare the performance of different algorithms. We consider the following 6 algorithms:  
\begin{itemize}
    \item The proposed RCG algorithm for CI (RCG-CI);
    \item Convex relaxation for CI (CVX-CI)\cite{7738555};
    \item Cross-entropy optimization for CI (CEO-CI)\cite{7738555};
    \item RCG algorithm for IR (RCG-IR)\cite{7811286};
    \item Gradient descent algorithm for IR (GD-IR)\cite{6451071};
    \item Cross-entropy optimization for IR (CEO-IR) \cite{6853373}.
\end{itemize}
Without loss of generality, we use QPSK modulation for all the approaches. We set $u = 1, \forall m$, which is a common assumption in related literature for the reason that the optimal $u$ is difficult to determine for IR methods\cite{6451071,6853373} while arbitrary $u$ can be accepted by CI methods \cite{7738555}. We also assume that $P_T = 1, N = 64$ for all the algorithms, and each entry of the channel $\mathbf{H}$ subjects to standard complex Gaussian distribution, i.e., $h_{n,m}\sim\mathcal{CN}\left(0,1\right), \forall n,\forall m$. For CEO methods, we use the same parameter configuration with \cite{7738555}, which is $T=1000$ (the number of iterations), $K = 500$ (the number of initialized random samples), $\rho = 0.05$ (quantile), $\alpha = 0.08$ (the smooth parameter). For GD-IR, the number of iterations is set as 50.
\\\indent While the analytic complexity per iteration of the most algorithms has already been given, we compare the overall complexity in terms of average execution time in Fig. 2 (a) since it is difficult to specify the complexity of the CVX-CI approach. The simulation is performed on an Intel Core i7-4790 CPU 32GB RAM computer with 3.6GHz. As expected, the RCG methods require least execution time to solve the problem while other methods need much more. Although the proposed RCG-CI algorithm is more complex than RCG-IR by each iteration, the total time needed is still comparable with the latter. More importantly, RCG-CI is robust to the increasing users because its complexity is mainly determined by the antenna number of the BS. 
\\\indent In Fig. 2 (b), we show the error performance of all 6 approaches in terms of SER with increased transmit signal-to-noise-ratio (SNR), where $M = 20$, $\text{SNR} = P_T/N_0$. Note that all the IR methods show negligible difference under the given parameter configuration, and all the CI methods outperform the IR methods thanks to the utilization of the MUI power. It is worth noting that the proposed RCG-CI has the best performance among all the 6 approaches with 2dB gain over IR methods, and 1dB gain against the CVX-CI algorithm. 
\\\indent We further consider the error performance with increased number of users in Fig. 2 (c), where the SNR is fixed at 8dB with the number of users ranging from 12 to 24. It can be observed that the SER becomes worse with the growth of the users due to the reduction of the Degrees of Freedom (DoFs). Once again, we see that the proposed RCG-CI achieves the lowest SER among all the approaches, and the CI methods achieve far better performance than IR methods, while the latter maintains an SER of $10^{-2}$ for all the users numbers. 

\section{Conclusion}
A low-complexity manifold optimization algorithm has been introduced to solve the CEP problem with the exploitation of the MUI power. By viewing the feasible region of the optimization as an oblique manifold, the proposed method can efficiently find a near-optimal solution using the Riemannnian conjugate gradient algorithm. Numerical results show that the proposed RCG-CI algorithm outperforms the existing 5 other approaches in terms of error performance, with a comparable complexity to the fastest RCG-IR algorithm. It is further shown that when the DoFs of the system are limited, the proposed RCG-CI still performs far better than other methods.


%




\ifCLASSOPTIONcaptionsoff
  \newpage
\fi



\bibliographystyle{IEEEtran}
\bibliography{IEEEabrv,Manifold_CEP_CI}

\begin{thebibliography}{10}
\providecommand{\url}[1]{#1}
\csname url@samestyle\endcsname
\providecommand{\newblock}{\relax}
\providecommand{\bibinfo}[2]{#2}
\providecommand{\BIBentrySTDinterwordspacing}{\spaceskip=0pt\relax}
\providecommand{\BIBentryALTinterwordstretchfactor}{4}
\providecommand{\BIBentryALTinterwordspacing}{\spaceskip=\fontdimen2\font plus
\BIBentryALTinterwordstretchfactor\fontdimen3\font minus
  \fontdimen4\font\relax}
\providecommand{\BIBforeignlanguage}[2]{{%
\expandafter\ifx\csname l@#1\endcsname\relax
\typeout{** WARNING: IEEEtran.bst: No hyphenation pattern has been}%
\typeout{** loaded for the language `#1'. Using the pattern for}%
\typeout{** the default language instead.}%
\else
\language=\csname l@#1\endcsname
\fi
#2}}
\providecommand{\BIBdecl}{\relax}
\BIBdecl

\bibitem{6736761}
E.~G. Larsson, O.~Edfors, F.~Tufvesson, and T.~L. Marzetta, ``Massive {MIMO}
  for next generation wireless systems,'' \emph{IEEE Communications Magazine},
  vol.~52, no.~2, pp. 186--195, February 2014.

\bibitem{5595728}
T.~L. Marzetta, ``Noncooperative cellular wireless with unlimited numbers of
  base station antennas,'' \emph{IEEE Transactions on Wireless Communications},
  vol.~9, no.~11, pp. 3590--3600, November 2010.

\bibitem{5978417}
V.~Mancuso and S.~Alouf, ``Reducing costs and pollution in cellular networks,''
  \emph{IEEE Communications Magazine}, vol.~49, no.~8, pp. 63--71, August 2011.

\bibitem{6297982}
S.~K. Mohammed and E.~G. Larsson, ``Single-user beamforming in large-scale
  {MISO} systems with per-antenna constant-envelope constraints: The doughnut
  channel,'' \emph{IEEE Transactions on Wireless Communications}, vol.~11,
  no.~11, pp. 3992--4005, November 2012.

\bibitem{6451071}
------, ``Per-antenna constant envelope precoding for large multi-user {MIMO}
  systems,'' \emph{IEEE Transactions on Communications}, vol.~61, no.~3, pp.
  1059--1071, March 2013.

\bibitem{6853373}
J.~C. Chen, C.~K. Wen, and K.~K. Wong, ``Improved constant envelope multiuser
  precoding for massive {MIMO} systems,'' \emph{IEEE Communications Letters},
  vol.~18, no.~8, pp. 1311--1314, Aug 2014.

\bibitem{7811286}
J.~C. Chen, ``Low-{PAPR} precoding design for massive multiuser {MIMO} systems
  via riemannian manifold optimization,'' \emph{IEEE Communications Letters},
  vol.~21, no.~4, pp. 945--948, April 2017.

\bibitem{7738555}
P.~V. Amadori and C.~Masouros, ``Constant envelope precoding by interference
  exploitation in phase shift keying-modulated multiuser transmission,''
  \emph{IEEE Transactions on Wireless Communications}, vol.~16, no.~1, pp.
  538--550, Jan 2017.

\bibitem{7103338}
C.~Masouros and G.~Zheng, ``Exploiting known interference as green signal power
  for downlink beamforming optimization,'' \emph{IEEE Transactions on Signal
  Processing}, vol.~63, no.~14, pp. 3628--3640, July 2015.

\bibitem{duan2013natural}
X.~Duan, H.~Sun, L.~Peng, and X.~Zhao, ``A natural gradient descent algorithm
  for the solution of discrete algebraic lyapunov equations based on the
  geodesic distance,'' \emph{Applied Mathematics and Computation}, vol. 219,
  no.~19, pp. 9899--9905, 2013.

\bibitem{li2016optimal}
C.~Li, E.~Zhang, L.~Jiu, and H.~Sun, ``Optimal control on special euclidean
  group via natural gradient algorithm,'' \emph{Science China Information
  Sciences}, vol.~59, no.~11, p. 112203, 2016.

\bibitem{4801492}
C.~Masouros and E.~Alsusa, ``Dynamic linear precoding for the exploitation of
  known interference in {MIMO} broadcast systems,'' \emph{IEEE Transactions on
  Wireless Communications}, vol.~8, no.~3, pp. 1396--1404, March 2009.

\bibitem{boumal2014manopt}
N.~Boumal, B.~Mishra, P.-A. Absil, and R.~Sepulchre, ``Manopt, a {M}atlab
  toolbox for optimization on manifolds,'' \emph{The Journal of Machine
  Learning Research}, vol.~15, no.~1, pp. 1455--1459, 2014.

\bibitem{petersen2006riemannian}
P.~Petersen, \emph{Riemannian geometry}.\hskip 1em plus 0.5em minus 0.4em\relax
  Springer, 1998, vol. 171.

\bibitem{boumal2014thesis}
N.~Boumal, ``Optimization and estimation on manifolds,'' Ph.D. dissertation,
  Universit\'e catholique de Louvain, February 2014.

\bibitem{absil2009optimization}
P.-A. Absil, R.~Mahony, and R.~Sepulchre, \emph{Optimization algorithms on
  matrix manifolds}.\hskip 1em plus 0.5em minus 0.4em\relax Princeton
  University Press, 2009.

\end{thebibliography}
\end{document}